\documentclass[reprint, aps, prx, longbibliography, superscriptaddress, twocolumn]{revtex4-2}
\usepackage[utf8]{inputenc}
\usepackage{braket}
\usepackage{amsmath}
\usepackage{amssymb}
\usepackage{float}
\usepackage{graphicx}
\usepackage{subcaption}
\usepackage{xcolor}
\usepackage{url}
\usepackage{bm}
\usepackage{enumerate}
\usepackage[final]{showlabels}

\newcommand{\ul}[1]{\underline{#1}}
\newcommand{\beq}{\begin{equation}}
\newcommand{\eeq}{\end{equation}}

\newcommand{\usigma}{\ul{\sigma}}
\newcommand{\utheta}{\ul{\theta}}

\long\def\ca#1\cb{} 

\begin{document}
\title{Autoregressive pairwise Graphical Models efficiently find ground state representations of stoquastic Hamiltonians}
\author{Yuchen Pang$^{*}$}
\thanks{ Contributed equally.}
\affiliation{University of Illinois at Urbana-Champaign, Department of Computer Science
Champaign, IL, USA}
\author{Abhijith Jayakumar$^{*}$} 
\affiliation{Theoretical Division, Los Alamos National Laboratory, Los Alamos, NM, USA}
\email{abhijithj@lanl.gov}
\author{Evan McKinney}
\affiliation{Department of Electrical and Computer Engineering, University of Pittsburgh, Pittsburgh, PA, USA}
\author{Carleton Coffrin}
\affiliation{Advanced Network Science Initiative, Los Alamos National Laboratory, Los Alamos, NM, USA}
\author{Marc Vuffray}
\affiliation{Theoretical Division, Los Alamos National Laboratory, Los Alamos, NM, USA}
\author{Andrey Y. Lokhov}
\email{lokhov@lanl.gov}
\affiliation{Theoretical Division, Los Alamos National Laboratory, Los Alamos, NM, USA}

\begin{abstract}
We introduce Autoregressive Graphical Models (AGMs) as an Ansatz for modeling the ground states of stoquastic Hamiltonians. Exact learning of these models for smaller systems show the dominance of the pairwise terms in the autoregressive decomposition, which informs our modeling choices when the Ansatz is used to find representations for ground states of larger systems. We find that simple AGMs with pairwise energy functions trained using first-order stochastic gradient methods often outperform more complex non-linear models trained using the more expensive stochastic reconfiguration method. We also test our models on Hamiltonians with frustration and observe that the simpler linear model used here shows faster convergence to the variational minimum in a resource-limited setting.
\end{abstract}

\maketitle
\begingroup\renewcommand\thefootnote{\textsection}
\footnotetext{Equal contribution}
\endgroup

\section{Introduction}

Approximating ground states of many-body quantum systems, in general, is known to become computationally intractable as the number of particles in the system increases. Nevertheless, computational methods currently remain the only feasible way to study the ground state properties of most quantum models. In the recent years, researchers have adopted and extended methods from machine learning to tackle this specific problem. The most  popular of these approaches involves using a neural network (NN) as a variational Ansatz for the ground state wave function. The variational parameters of the net are then optimized using the well-known Variational Monte Carlo (VMC) method \cite{carleo2017solving, selke1981finite,mcmillan1965ground, ceperley1977monte}. Unlike tensor network based methods, neural network quantum states have been observed to be capable of efficiently representing states that have a volume law entanglement\cite{deng2017quantum}. This makes these neural nets an effective, complementary approach to tensor network methods for the problem of finding ground states. 

The success of this type of methods depend crucially on choosing the right neural net to represent the state. The network must be able to approximate the true ground state for some choice of the variational parameters. While the universal approximation theorem for neural nets guarantees that this condition will be satisfied if we take a large enough network \cite{westerhout2020generalization}, it is unclear if an effective representation for ground states can be learned using a simpler Ansatz that could be advantageous in data-limited setting. The use of neural nets as a variational Ansatz was pioneered by Carleo and Troyer \cite{carleo2017solving}, who used a Restricted Boltzmann machine architecture to find the ground states of spin Hamiltonians using VMC. Since then there have been a vast number of works that used the framework proposed in \cite{carleo2017solving}, with different NN architectures on  a variety of related problems \cite{liang2021hybrid,choo2019two, choo2018symmetries, choo2020fermionic, sharir2020deep, lange2024architectures}. Notably, many NN architectures have been introduced which implicitly respect symmetries of the model being solved \cite{vieijra2020restricted, roth2021group}, like the use of CNNs for models with translational invariance\cite{choo2019two}.

Since no single Ansatz can be expected to work well for all models and for all regimes, it is important to look at these methods more carefully and understand their drawbacks \cite{schollwock2011density,verstraete2008matrix}. For instance, an MPS Ansatz for a state can be expected to perform poorly near a quantum critical point where long-range correlations are be present. NNs are not easy to understand in this regard as they are essentially black boxes. Given a specific NN architecture it would be difficult to assess why it performs better in one case as opposed to another. Ideally, there must exist some way to choose an Ansatz without performing a painstaking search over architectures. We would like this search to be replaced by some simpler procedure that would guide our intuition.

Recently, classical energy-based representations have been proposed to model the ground states of quantum multi-body Hamiltonians \cite{jayakumar2023learning}. This classical Gibbs representation is tightly connected to the concept of undirected graphical models. In these models, a graph encodes conditional independence relationships in the following way: any single variable in the model is conditionally independent of the rest of the graph given its neighbors. This conditional dependency structure thus allows us to represent these distributions using very few parameters \cite{mezard2009information,clifford1990markov}. This is a desirable property, because physical systems often display such sparsity properties at the level of the Hamiltonian while this sparsity might not be evident when looking at the probability distribution representing these systems. Moreover, these classical thermal distributions are explicitly learnable using low-complexity algorithms \cite{vuffray2016interaction, lokhov2018optimal, vuffray2019efficient, abhijith2020learning, jayakumar2024discrete}, and the emerging effective temperature of the learned energy function serves as a metric that quantifies the hardness of learning of different classes of quantum states \cite{jayakumar2023learning}. In energy-based modeling, energy functions can be specified by a small number of parameters using an explicit low-degree polynomial or a generic parametric family such as neural nets. These choices naturally provide a trade-off between simplicity and generality of the representation.

One of the main drawbacks of using Graphical Models in representing probability distributions is the lack of exact sampling algorithms that can efficiently produce uncorrelated samples from the model. Learned energy-based representations directly produce conditional probabilities that can be directly used in Markov chain Monte Carlo methods \cite{jayakumar2023learning}. However, commonly used Markov chain methods can often take an exponentially long time to produce independent samples. This issue is  especially severe for models with frustration where the Markov chain can get stuck in meta-stable states far from the stationary distribution \cite{mackenzie1982lack, krauth2006statistical}. On the other hand autoregressive models \cite{sharir2020deep, wu2019solving}, provide a way to model distributions such that independent samples can be efficiently produced. But these models are generally constructed using a non-local marginalization procedure that usually doesn't respect the conditional independence structure of the distribution.

In this work, we build on the energy-based modeling approach and propose a new variational Ansatz for stoquastic Hamiltonians that we call Autoregressive Graphical Model (AGM). With the AGM model, we will combine both the representation power of GMs and the sampling efficiency of the autoregressive models. This representation is flexible enough to incorporate non-linear parametrizations like neural networks \cite{jayakumar2023learning}, but the focus of this work will be on explicit polynomial modeling of the energy function that provides interpretability. We use an exact learning procedure to find the true autoregressive model corresponding to the ground states of smaller systems. The information obtained from these smaller experiments can then be used to understand the applicability of this Ansatz for larger systems. We find from exact learning that autoregressive model representing ground states of different models are dominated by pairwise interactions. Subsequently, we show that a simple pairwise energy Ansatz used in the autoregressive framework is sufficient to capture the ground state properties of a variety of stoquastic quantum models including those with frustration.

Interestingly, the resulting emerging model is reminiscent of the well-studied Jastrow wavefunction Ansatz which has been used to model correlations in a quantum system in a principled manner \cite{jastrow1955many,mahajan2019symmetry}. For the stoquastic spin systems, the Jastrow Ansatz exactly corresponds to a graphical model representation for the \emph{full} wave function. Instead, in our approach, a \emph{local} pairwise energy function Ansatz is used to model each of the \emph{conditional distributions} in the autoregressive decomposition. We find that these simple models trained using only first-order stochastic optimization methods like ADAM \cite{DBLP:journals/corr/KingmaB14} are able to perform comparably in certain regions of the zero-temperature phase space to more complex architectures such as neural net quantum states (implemented through the state-of-the-art toolbox NetKet \cite{vicentini2021netket}) of similar size trained using second-order optimization methods when compared in the same time frame. For non-frustrated models, we also provide benchmarks produced by Tensor Network methods. Overall, our results suggest a further exploration of energy-based modeling as variational Ansatz for quantum systems. 

\section{Problem formulation}

The problem of finding the ground state of any quantum Hamiltonian can be cast
as a variational problem. Given  a $n$-qubit Hamiltonian $H$, the following variational problem can be solved to get the ground state energy,
\begin{equation}\label{eq:var_full}
E_{g} =  { \displaystyle \min_{\ket{\psi} \in \mathbb{C}^{2^n}, || \: \ket{\psi} ||_2 = 1 }}   \braket{\psi | H | \psi}
\end{equation}
We call the matrix element, $\braket{\psi | H | \psi}$, the energy of the
state $\ket{\psi}$.
A ground state, $\ket{\psi_{g}}$ is the solution to the above minimization problem,
\begin{equation}
  \braket{\psi_g | H | \psi_g} = E_g.
\end{equation}
The minimization in \eqref{eq:var_full} is over a Hilbert space whose dimension is exponentially large in the number of spins. This makes the exact minimization of \eqref{eq:var_full} intractable for all but small quantum systems.  

Hamiltonians with non-positive off-diagonal entries are known as
\emph{stoquastic} or sign-problem free. The ground states of these Hamiltonians
can be chosen to have non-negative components in the computational basis $(~~\left\{
\ket{\usigma}~|~\usigma \in \{1,-1\}^n \right\}~~)$ \cite{bravyi2014monte}.

For such Hamiltonians the variational problem can be reduced to one over
classical probability distributions over $n$ spins. To make the minimization
tractable using a Monte-Carlo approach, first we define a family of quantum states from a  family of
probability distributions parameterized by a set of real variables ($\utheta$),
$$ \ket{P(\utheta)} = \sum_{\usigma}  \sqrt{P(\usigma;\utheta)} \ket{\usigma}. $$
This leads to an approximate variational problem for the ground state.
The variational energy, $E(\utheta)~ \equiv ~ \braket{P(\utheta)| H | P(\utheta)} $, can be minimized over the space of the $\utheta$ parameters to get an approximate solution to the problem in \eqref{eq:var_full}.
The quality of the approximation depends on how well the parametric family can
approximate the ground state.

The energy $E(\theta)$ can be written as an expectation value of a random
variable. The energy estimate  written in this form is,
\ca
\begin{align}
  E(\utheta) &=  \sum_{\usigma, \usigma^\prime} H_{\usigma, \usigma^\prime}~ \sqrt{P(\usigma;\utheta) P(\usigma^\prime; \utheta)}\\
    &= \sum_{\usigma, \usigma^\prime} P(\usigma;\utheta)  H_{\usigma, \usigma^\prime}~  \sqrt{ \frac{P(\usigma^\prime;\utheta)} {P(\usigma; \utheta)}} 
     = \underset{\usigma \sim  P(;\utheta)}{\mathbb{E}}~~ E_{loc}(\usigma; \utheta)
\end{align}
\cb

\begin{align}
  E(\utheta) = \sum_{\usigma, \usigma^\prime} P(\usigma;\utheta)  H_{\usigma, \usigma^\prime}~  \sqrt{ \frac{P(\usigma^\prime;\utheta)} {P(\usigma; \utheta)}} 
     = \underset{\usigma \sim  P(;\utheta)}{\mathbb{E}}~~ E_{loc}(\usigma; \utheta)
\end{align}

Here $E_{loc}(\usigma; \utheta) \equiv \sum_{\usigma^\prime}   H_{\usigma, \usigma^\prime}~  \sqrt{ \frac{P(\usigma^\prime;\utheta)} {P(\usigma; \utheta)}} $ , is known in literature as the \emph{local energy}.
The local energy for any spin-configuration can be efficiently computed given that $H$ is sparse. and this condition is satisfied for all local Hamiltonains defined on a graph with bounded degree.
\subsection*{Monte-Carlo approximation}

 Given the ability to compute the local energy and given i.i.d samples $\{ \usigma^{(1)}
, \ldots, \usigma^{(N_{s})}\}$ from $P(\utheta)$, the we can compute an unbiased
estimate for $E(\utheta)$,
\begin{equation}\label{eq:en_est}
  \hat{E}(\utheta) = \frac{1}{N_s} \sum_{t}  E_{loc}(\usigma^{(t)}; \utheta).
\end{equation}

Similarly an unbiased estimate for the gradient of $E(\utheta)$ can also be computed
\begin{align}
  \nabla _{\utheta} E(\utheta) &=  \underset{\usigma \sim  P(;\utheta)}{\mathbb{E}}~~  E_{loc}(\usigma; \utheta) \nabla_{\utheta} ~\ln ( P(\usigma; \utheta)),\\
                               &\approx \frac{1}{N_s} \sum_{t}  E_{loc}(\usigma^{(t)}; \utheta)  \nabla_{\utheta}\ln ( P(\usigma^{(t)}; \utheta)) \label{eq:grad_est}
\end{align}

It is crucial to have high quality samples drawn from $P(\underline{\sigma}; \utheta)$ to compute these estimates. So it is desirable to have a model that allows for exact sampling without Markov chain methods.

\subsection*{Autoregressive Graphical Models}

Following \cite{jayakumar2023learning}, we propose to use energy-based modeling to represent $P(\underline{\sigma}; \utheta)$. However, our aim here is to construct an Ansatz that will retain the advantages of the graphical model approach while still allowing for exact sampling. To achieve this, first we factorize the Ansatz distribution using an autoregressive form,
$$ P(\underline{\sigma}; \utheta) =  \prod_i P(\sigma_i | \sigma_{>i}; \utheta^{(i)})$$
Then we directly parametrize the energy associated with each conditional using a function having at most pairwise interactions,
\ca
\begin{align}\label{eq:AGM_pairwise}
P(\sigma_i|\sigma_{>i}; \utheta^{(i)}) &= \frac{\exp \left( \sigma_i ( \theta^{(i)}_{0} +  \sum_{j >i} \theta^{(i)}_{j} \sigma_j ) \right)}{2 \cosh \left( \sigma_i ( \theta^{(i)}_{i} +  \sum_{j >i} \theta^{(i)}_{j} \sigma_j ) \right) } \\
 &:=   S(\sigma_i, \sigma_{>i}; \utheta^{(i)}).
\end{align}
\cb
\begin{align}\label{eq:AGM_pairwise}
P(\sigma_i|\sigma_{>i}; \utheta^{(i)}) &= \frac{1}{1 +  \exp \left (-2\sigma_i ( \theta^{(i)}_{i} +  \sum_{j >i} \theta^{(i)}_{j} \sigma_j ) \right) } \\
 &:=   S(\sigma_i, \sigma_{>i}; \utheta^{(i)}).
\end{align}

We refer to this Ansatz as to the Autoregressive Graphical Model. It allows for exact sampling form the distribution and each $P(\usigma;\utheta)$ can be computed as well. Moreover, for a local Hamiltonian, we can estimate the variational energy and its gradients in $O(n^2)$ time. Higher-order terms can be added to the Ansatz, thus increasing the expressivity of the model at the cost of its complexity. But in this work we show that using an AGM with up to pairwise interactions already give good results for larger systems, as explained next.

  \begin{figure*}
    \centering
    \begin{subfigure}[b]{0.32\textwidth}
    \includegraphics[height=0.7\textwidth]{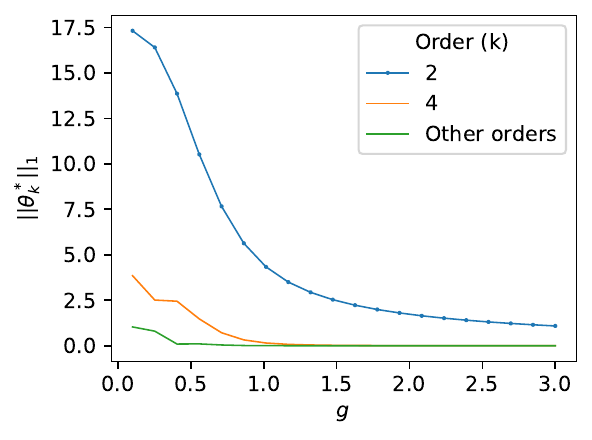}
   \caption{}
\end{subfigure}    
\begin{subfigure}[b]{0.32\textwidth}
    \centering  
    \includegraphics[height=0.7\textwidth]{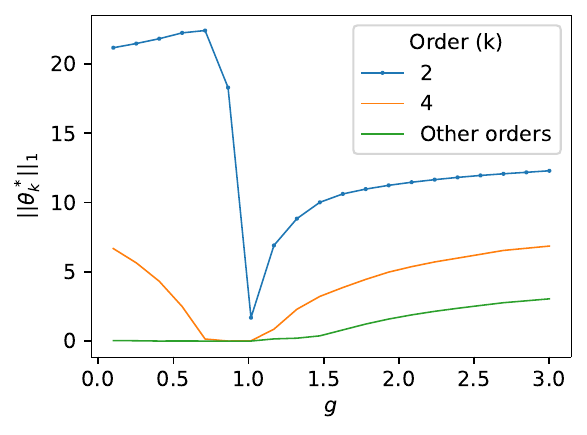}
  \caption{}
\end{subfigure}
\begin{subfigure}[b]{0.32\textwidth}
    \centering  
    \includegraphics[height=0.7\textwidth]{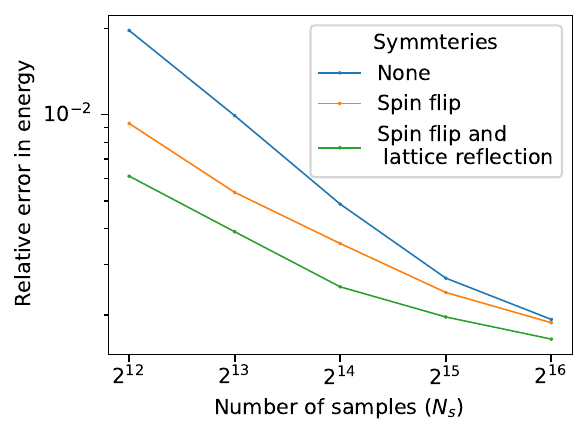}
  \caption{}
      \label{fig:1Dsym}
\end{subfigure}
\caption{{\bf Results for 1D models.} The magnitude of the largest term at each order is presented for the autoregressive conditionals learned from the ground state of an (a) Ferromagnetic TIM (b) Ferromagnetic XXZ Model. Both models are 1D chains on $7$ spins. For the Heisenberg model, we break the ground state degeneracy using a small $Z$ field which preserved the stoquasticity of the model. (c) Effect of increasing the number of samples and using prior information about symmetries in the linear GM Ansatz. The model used here is a 1D $H_{TIM}$ Hamiltonian on $100$ spins with open boundary conditions.}
\label{fig:Order_1D}
\end{figure*}

\section{Results}

\subsection{Applicability of the pairwise Ansatz}

In general, the affine function inside the sigmoid in \eqref{eq:AGM_pairwise} can be replaced with a general polynomial on $n-i$ variables. This would be the most general model and would be able to represent any state exactly at the cost of exponential computational complexity.  While this is impractical for larger systems, learning this complete model is tractable for systems with a few spins. For a given family of models, we can study the non-zero terms in the general polynomial and this can inform the choice of Ansatz for larger systems in the same family. 

To find the exact AGM representation for smaller systems, we use a method called \textit{Interaction Screening} \cite{vuffray2016interaction,lokhov2018optimal, vuffray2019efficient, abhijith2020learning} which is an efficient machine learning algorithm for estimating the energy function of a graphical model given samples from it. For the case of smaller systems, we can make this method exact by using the most general representation for the AGM as mentioned above and by artificially taking the infinite sample limit. We use this method as it gives a convex optimization method for finding the AGM and hence we can guarantee global convergence. This is not possible using the variational method as convexity of the variational energy is not guaranteed even for the simple pairwise autoregressive Ansatz.

Interaction screening  learns the exact form of the autoregressive representation by minimizing the following objective,
\begin{equation}\label{eq:exact}
   \utheta^{*(i)} = \underset{\utheta}{\text{argmin}}\sum_{\usigma \in \{1,-1\}^n} |\braket{\usigma|\psi_g}|^2 \exp(-\sigma_i  F_i(\sigma_{>i});\utheta^{(i)}).
\end{equation}
Where $F_i(\sigma_{>i};\utheta^{(i)}): \mathbb{R}^{n-i} \rightarrow \mathbb{R}$ is most general polynomial function on $n-i$ binary variables parameterized as $\theta_0^{(i)} + \sum_{j>i} \theta^{(i)}_{j}\sigma_j + \sum_{j>i, k>j} \theta^{(i)}_{j,k} \sigma_j \sigma_k + \ldots$ 

Now, from the consistency of the Interaction screening estimator ( see \cite{abhijith2020learning, vuffray2019efficient} ) the solution of the optimization problem will uniquely represent the ground state in  the autoregressive form,
\begin{align}
|\braket{\usigma | \psi_g}|^2 &=  \prod_i  P_g(\sigma_i | \sigma_{>i}), \\
P_g(\sigma_i|\sigma_{>i}) &= \frac{1}{ 1+ \exp \left(-2\sigma_i F(\sigma_{>i}, \utheta^{*(i)})  \right) }
\end{align}

For a system of few spins, the exact evaluation of the objective in \eqref{eq:exact} is tractable.  We can compute the true ground state $\ket{\psi_g}$ by brute-force diagonalization and solve the convex optimization problem outlined in \eqref{eq:exact} to find the complete representation of the ground state in the autoregressive form.

In \figurename \ref{fig:Order_1D}, we use this method to find the exact AGM representation for seven spins in a 1D chain interacting according to the Transverse Ising Model (TIM) and XXZ Hamiltonians. To study the relative importance of terms at each order we look at the sum of absolute values in the solution of \eqref{eq:exact} for every order ($||\theta^*_k||_1$ for order $k$). The results of these experiments show that AGMs learned from these models are dominated by pairwise terms which implies the applicability of the pairwise autoregressive model as a variational Ansatz for finding these ground states. 
\subsection{Models without frustration}
Supported by the dominance of pairwise terms observed in \figurename \ref{fig:Order_1D} we now use the AGM Ansatz to find the ground states of systems of larger size. Here we look at two classes of  models without frustration, namely ferromagnetic TIM models and anti-ferromagnetic  XXZ models.
\begin{figure}[h]
    \includegraphics[width=0.45\textwidth]{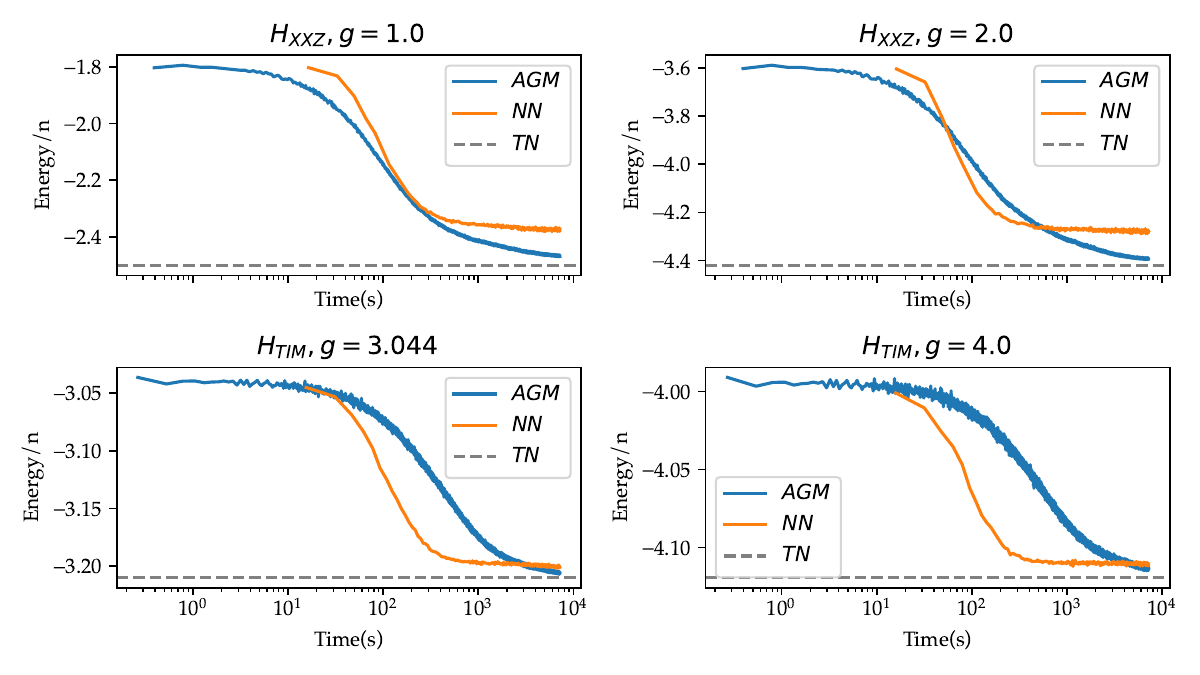}
    \caption{{\bf Results for 2D models without frustration.} Variational energy of AGM Ansatz vs time while trained on a $10\times 10$ Ferromagnetic $H_{TIM}$ and Antiferromagnetic $H_{XXZ}$ models with open boundary conditions. We see that the linear AGM Ansatz has comparable performance to the non-linear neural net Ansatz when trained with the same number of samples. The non-linear model is chosen such that it has the same number of trainable parameters as the linear GM model.}
    \label{fig:non_frust_2D}
\end{figure}

\begin{equation}
    H_{TIM} = -\sum_{<i,j>} Z_i Z_j - g \sum_{i} X_i 
\end{equation}

To begin with we try to learn ground state representations for the simpler case of this model on a one-dimensional lattice of spins. Being a one-dimensional model, an almost exact representation of the ground state can be found using DMRG \cite{white1993density, itensor}. For 1D models, DMRG gives excellent results owing to the ability of the matrix-product Ansatz implicit in DMRG methods being able to approximate the ground state using a number of parameters tractable in the problem size. 
We use the DMRG result as a benchmark to study the performance of our method with a linear energy function. We find that two different approaches can be taken to reduce the error in the ground state. The first one is simply by increasing the number of samples $N_s$ and the second is by incorporating prior information about the symmetries of the Hamiltonian into the model.

The 1D Hamiltonian  with open boundaries has two obvious symmetries that we can exploit. The Hamiltonian invariant under a global spinflip (the $X$ operator applied to every site) and also invariant under a reflection of the lattice about its mid-point (swapping the $i$-th qubit with the $n-i +1$ qubit). Both of these correspond to a two-element symmetry group and can be easily implemented by appropriately averaging the Ansatz.
The effect of both these approaches is shown in \figurename \ref{fig:1Dsym}. Here we see that increasing the number of samples with the linear energy Ansatz consistently reduces the error in the ground state energy. Symmetrizing the anstaz can help reduce the error when the number of samples is low but this advantage vanishes when the number of samples becomes very large. This implies that our model is able to discover the symmetries implicitly given the samples. While both these approaches increase the computational cost of training, including symmetries mainly affects the running time of the algorithm and not its memory consumption. While increasing the number of samples affects both computational time and memory usage of the algorithm. 

Next, we use AGMs on 2D stoquastic models. For these experiments, we use ferromagnetic TIM models and anti-ferromagnetic XXZ models,
\begin{equation}
    H_{XXZ} = \sum_{\langle i,j \rangle} Z_iZ_j - g(X_iX_j + Y_iY_j).
\end{equation}
We compare the AGM Ansatz with a Neural network quantum state (NN) \cite{vicentini2021netket} with the same number of trainable parameters, and a Tensor network (TN) based approach that uses imaginary time evolution of to find the ground states \cite{cirac2009renormalization,jiang2008accurate,pang2020efficient}. The reported TN energy in \figurename \ref{fig:non_frust_2D} is obtained after experimenting with different update rules and bond dimensions (see Appendix \ref{app:methods}). For the NN training, we used a fixed set of hyper-parameters that were observed to give good results. For AGM we used a hyperparameter optimization routine that used only $Ns = 2^8$ samples. The hyperparameters obtained from this low-cost simulation were seen to give good results even for the case of more samples.  Note that we use the NN  results as a baseline to study the  behavior of the AGM Ansatz. The performance of the NN method can presumably be increased further by using more intense hyperoptimization.  

In the interest of studying the practicality of these methods, we set a maximum to compute time of approximately 2 hrs for all the methods on the same hardware. The TN method gives lower energies than both the machine learning-based approaches. But this is not a variational method, and moreover, attempts to improve the accuracy of the TN algorithm are seen to drastically increase its run time, due to the complexities of working with the PEPS Ansatz used in this method.

Similar to the 1D case, we found that increasing the number of samples improved the variational energy of AGM. For NN, increasing the number of training parameters was seen to be the more effective strategy for improving the variational energy. Our experiments with both $H_{TIM}$ and $H_{XXZ}$, with varying values of $g$ showed that AGM is a better Ansatz than NNQS (as implemented in NetKet toolbox \cite{vicentini2021netket} and using preconditioned gradient descent method known as Stochastic Reconfiguration, see Appendix \ref{app:methods} for details) in the ``quantum dominated" part of the phase space, i.e when the off-diagonal elements in the Hamiltonian are dominant. The complete results of this comparison can be found in Appendix \ref{app:TIM_results}.

\subsection{Models with frustation}
MCMC-based  algorithms perform poorly when used to study systems with a complicated energy landscape. One way to construct such a system is by introducing frustration. Frustration is caused by competing interactions that cannot be simultaneously satisfied, leading to a highly degenerate ground state \cite{saul19942d}. It is well known that these models can cause Markov chain-based samplers to perform poorly due to a complex energy landscape that limits the ergodicity of these methods, particularly in the limit of large system sizes \cite{mezard2009information}.
This phenomenon leads to non-trivial resource trade-offs while considering ML-based models to represent quantum states. Using a variational Ansatz from which exact sampling is feasible gives in principle a way to overcome these limitations. But if we run the VMC algorithm for a long enough time, we expect the Markov chain to mix and give comparable or better performance to exactly samplable models. Hence in this context, the advantage of using models like AGM can be expected in a resource-limited setting, i.e. where we are constrained by time and memory. 

\begin{figure}
    \centering
    \begin{subfigure}[b]{0.225\textwidth}
    \includegraphics[width=\textwidth]{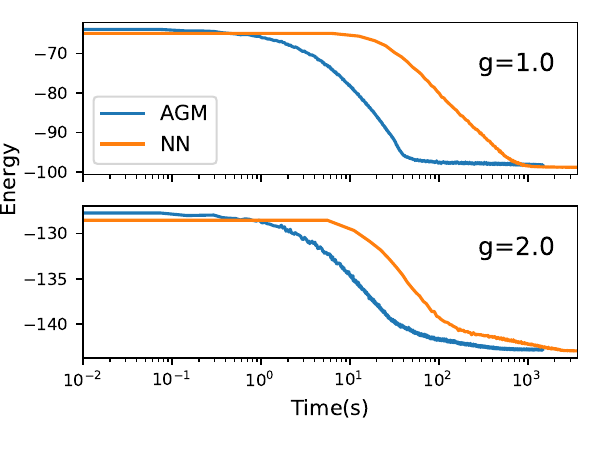}
   \caption{}
\end{subfigure}    
\begin{subfigure}[b]{0.225\textwidth}
    \centering  
    \includegraphics[width=\textwidth]{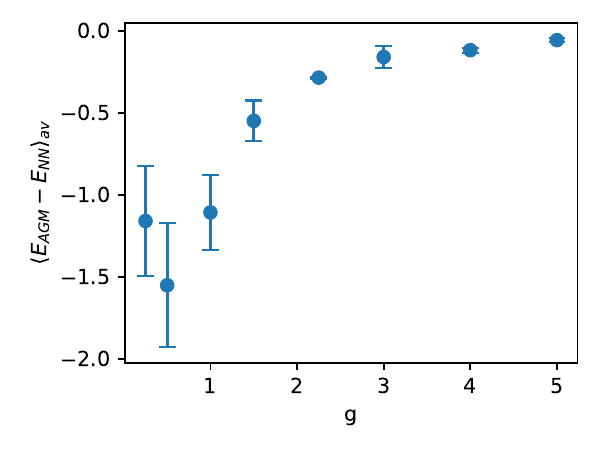}
  \caption{}
\end{subfigure}
\caption{{\bf Results for a 2D model with frustration.} VQMC results for finding the ground state of a disordered model $H_{D-TIM}.$ (a) The variational energy as a function of times for a specific disorder realization  of $8 \times 8$ model with $N_s = 2^{12}.$ We see that for smaller $g$ the AGM converges faster. (b) The difference in variational energy of AGM and NN models of the same size averaged over disorder for $6 \times 6$ models with $N_s = 2^{12}$ after running the VQMC algorithm for $200s$. Both plots show that AGM models give better variational energy in a resource-limited setting, particularly for the lower values of $g.$}
\label{fig:D_TIM}
\end{figure}

To test this we compare the performance of our method on a 2-D Transverse Ising model with disordered couplings \eqref{eq:D_TIM}. 
\begin{eqnarray}\label{eq:D_TIM}
    H_{D-TIM} = \sum_{<i,j>} J_{ij} Z_i Z_j - g \sum_{i} X_i ~~~ \\
    J_{ij} \sim \text{Unif(\{1, -1\})} \nonumber
\end{eqnarray}
For $g=0$ this is the well-known $\pm J$ spin glass model,  with is known to have a highly degenerate ground state \cite{blackman1991gauge, saul19942d}.  In the presence of the transverse field, this model is expected to have have a frustrated ground state for smaller values of $g$ \cite{dos1985transverse, chakrabarti1981critical}. 

\begin{figure}[bh]
    \centering
    \begin{subfigure}[t]{0.225\textwidth}
    \includegraphics[scale=0.16]{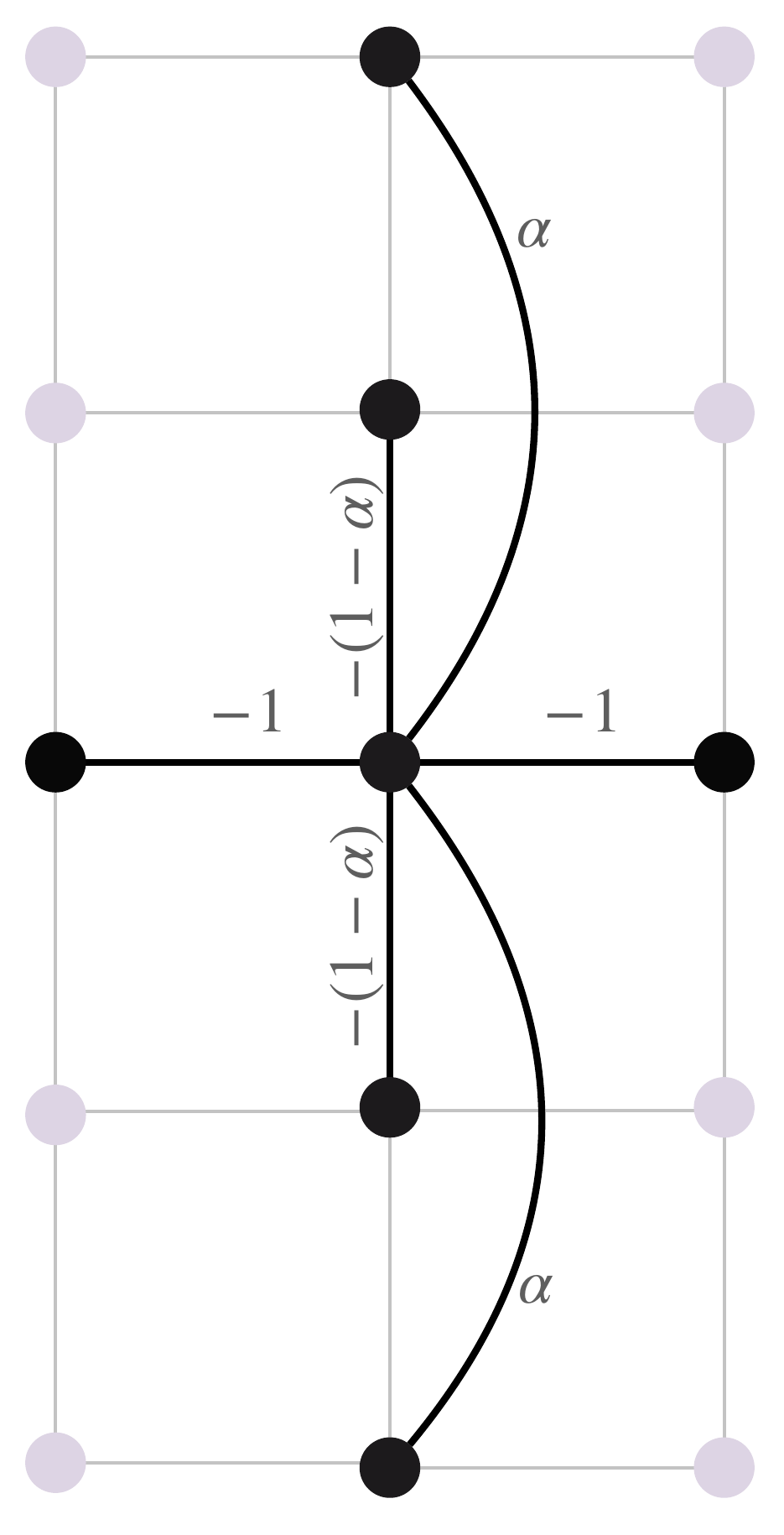}
    \caption{}\label{fig:annni_sketch}
    
    \end{subfigure}
    \begin{subfigure}[t]{0.225\textwidth}
    \includegraphics[width=\textwidth]{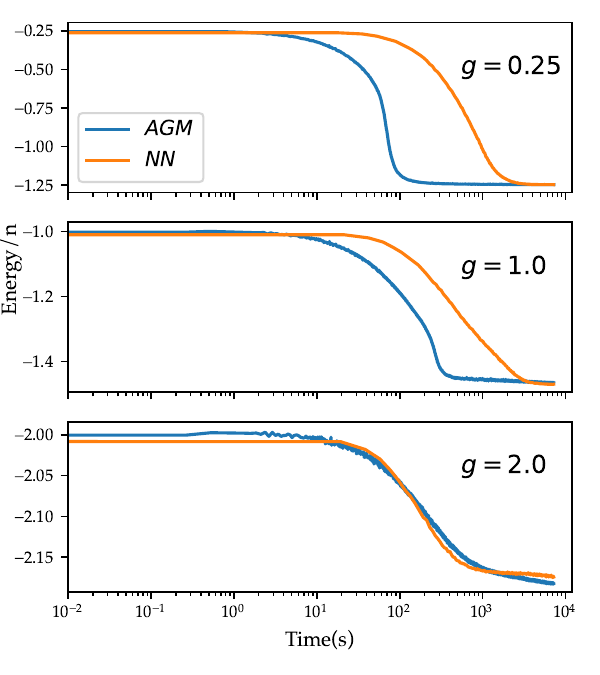}
    \caption{}\label{fig:annni}
    \end{subfigure}
    \caption{{\bf Results for the 2D Axial Next Nearest Neighbor Ising model with ground-state degeneracy.} (a) Two body terms in the 2D ANNI model. Notice the competing interactions induced by the \emph{n.n}  and \emph{n.n.n} couplings.  At $\alpha=1/3$, these lead to a highly degenerate ground state \cite{selke1980two}. (b) Variational energy as a function of time for the $10 \times 10$ ANNNI model with a transverse field ($N_s = 2^{12}$). We see that the AGM Ansatz is well suited in the small $g$ regime.}
\end{figure}

The results of VQMC simulations with this model are shown in \figurename \ref{fig:D_TIM}. We see that the AGM model converges faster than the NN model for this Hamiltonian for smaller values of $g.$ But the NN overtakes AGM when it is run for a longer time, which allows that Markov chain to produce better quality samples. To test this assertion, in  \figurename \ref{fig:D_TIM} (a), we look at the variational energy as a function of time for a single disorder realization of the disordered model on an $8 \times 8$ lattice. We observe that AGM converges faster, while the NN model takes considerably more time to converge. We find that this effect is much more pronounced for smaller values of $g$. Eventually, the NN is able to mix and give better variational energy, but the time taken for this is at least an order of magnitude larger for $g=1.$  To make sure that this is not a feature of a particular disorder realization, we perform a disorder-averaged experiment on $6 \times 6 $ lattices. In  \figurename \ref{fig:D_TIM} (b), we systematically study the average difference in variational energies found by the two methods in a setting where both the algorithms are run on the same hardware and the total run time is limited. We see that AGM outperforms NN in this setting, more so for smaller values of $g.$ 
The results in \figurename \ref{fig:D_TIM} suggest that the MCMC approach to VQMC struggles for models with frustration and the exact sampling approach is ideal in this regime. To test this further we look at the Axial Next Nearest Neighbor Ising (ANNNI) model on a 2D lattice with a transverse field,
\begin{eqnarray}
    H_{ANNNI} = &\sum_{x,y} -(1-\alpha) \, Z_{x,y} Z_{x,y\pm 1} +  \alpha \, Z_{x,y} Z_{x,y\pm 2}  - \nonumber \\
    & \,  Z_{x,y} Z_{x\pm1,y}  - g \sum_{x,y} X_{x,y}.  \label{eq:ANNNI}
\end{eqnarray}
This model has nearest-neighbor ferromagnetic interactions competing with next-nearest neighbor anti-ferro magnetic interactions along the $y-$axis (See \figurename ~\ref{fig:annni_sketch}). Notice that  these anti-ferro magnetic interactions are only present in the horizontal directions. Previous works on the classical variant of the  Hamiltonian in \eqref{eq:ANNNI} $(g=0)$ has shown that this model has a highly degenerate ground state  $\alpha = 1/3$, and for an $L\times L$ lattice the number of groundstates of the classical model is known to scale as $O(2^{cL})$ \cite{selke1980two, selke1981finite, redner1981one} for a constant $c$. 
This implies that for lower values of $g$, $H_{ANNNI}$ should be a hard model for MCMC-based VQMC methods compared to the exact sampling method. While for larger values of $g$, this gap should vanish due to the ground state being closer to the uniform superposition state. The results of this experiment are for $H_{ANNNI}$ on a $10 \times 10$ lattice is given in \figurename \ref{fig:annni}. The training here is done by drawing $2^{12}$ samples from the variational model. As expected, we see that got lower values of $g$ the AGM model converges much faster than the NN model.

\section{Conclusions}
In this work, we have introduced a simple variational model for learning ground states for stoquastic Hamiltonians. This model uses an autoregressive factorization of the ground state with an energy based representation of the conditionals. While the energy function in this model can be factorized using any function family (including neural nets), we find that the simple pairwise model is sufficient to capture the most dominant terms. We find that the pairwise model is capable of finding ground state energies of large stoquastic models and that its accuracy can be systematically improved by increasing the number of samples and by incorporating known symmetries. For the case of frustrated systems, we see that this exact sampling approach shows an advantage over an MCMC based approach, in the short time/low sample regimes.

Overall  the AGM Ansatz is especially seen to be useful in a resource-limited setting. It would be interesting to see if the solutions found by these models can be used to inform more resource-intense VQMC simulations using techniques from transfer learning \cite{yosinski2014transferable, bengio2012deep}. Possible avenues for future exploration involve increasing the expressivity of the AGM Ansatz by adding more complex terms beyond pairwise in the energy function \cite{jayakumar2023learning}, as well as incorporating a principled way of choosing the order of variables producing the best autoregressive decompositions.

\section*{Acknowledgements}
The authors acknowledge support from the Laboratory Directed Research and Development program of Los Alamos National Laboratory (LANL) under Projects No. 20240032DR and No. 20240245ER, and from the U.S. Department of Energy Office of Science Advanced Scientific Computing Research Program. EM was supported in part by the Quantum Computing Summer School Fellowship at LANL.




\appendix


\section{Methods and implementation details}
\label{app:methods}

\subsection{Tensor networks}

Tensor networks are one of the state-of-arts techniques to solve the ground states of many-body quantum systems~\cite{orus2014practical}.
In our benchmarks, we adopt the \textit{Quimb} library~\cite{gray2018quimb} and the \textit{ITensor} library~\cite{itensor} as reference implementations of tensor network algorithms.

The algorithm used for 1D TIM is the well-known density matrix renormalization group (DMRG) algorithm~\cite{white1993density} with matrix product states (MPS)~\cite{ostlund1995thermodynamic}. In particular, the reference ground state energy in Figure~\ref{fig:1Dsym} is computed with the DMRG algorithm with maximal bond dimension of $100$ and a cutoff error of $1\mathrm{E}{-}10$. Thanks to the effectiveness of the DMRG algorithm, the calculated ground state energy is almost exact.

The algorithm used for 2D TIM and antiferromagnetic Heisenberg model is imaginary time evolution (ITE) on the projected entangled pair states (PEPS)~\cite{verstraete2004renormalization} with the simple update~\cite{jiang2008accurate} and full update algorithm~\cite{jordan2008classical,lubasch2014algorithms}. Unlike MPS, these algorithms on PEPS generally have much higher time and space complexity. Therefore, we have to make a trade-off between accuracy and computational cost. Due to the limitation of time and resource, we choose to run 2D PEPS algorithms on a computer with two Broadwell-EP 12-core Xeon CPU and 256GiB memory, and the time used for each run is limited to roughly 10 hours.

To give some details of the PEPS algorithms we used, the PEPS entries are initialized with normal distributions. The PEPS are first evolved in imaginary time by $N_{SU}$ steps each for time steps $\mathbf\tau_{SU}=(\tau_1,\tau_2,\dots)$ using the simple update algorithm. Then optionally, the PEPS are evolved in imaginary time by $N_{FU}$ steps each for time steps $\mathbf\tau_{FU}=(\tau_1',\tau_2',\dots)$ using the full update algorithm, which does a better job in approximating ITE but has much higher cost. Besides $N_{SU},N_{FU},\mathbf\tau_{SU},\mathbf\tau_{FU}$, we also need to decide the bond dimension $D$ of PEPS and the contraction bond dimension $\chi$ for calculating energies and performing full updates. In particular, we set $\chi=4D^2$ since $\chi$ is often empirically chosen to be $\Theta(D^2)$. Note that the time and space complexity of PEPS contraction scale like $O(D^6)$ and $O(D^8)$ respectively~\cite{pang2020efficient}, so only small $D$ are tractable. 

In this work, we experiment with the following settings of the PEPS algorithms according to our hardware and time limit, and pick the best result.
\begin{enumerate}[(1)]
\item $D=4$,
    $N_{SU}=100$, $\mathbf\tau_{SU}=(0.3, 0.1, 0.03, 0.01, 0.003, 0.001)$;
\item $D=5$,
    $N_{SU}=100$, $\mathbf\tau_{SU}=(0.3, 0.1, 0.03, 0.01, 0.003, 0.001)$;
\item $D=6$,
    $N_{SU}=100$, $\mathbf\tau_{SU}=(0.3, 0.1, 0.03, 0.01, 0.003, 0.001)$;
\item $D=2$,
    $N_{SU}=100$, $\mathbf\tau_{SU}=(0.1, 0.03, 0.01)$,
    $N_{FU}=50$, $\mathbf\tau_{FU}=(0.3, 0.1, 0.03, 0.01)$;
\item $D=3$,
    $N_{SU}=100$, $\mathbf\tau_{SU}=(0.1, 0.03, 0.01)$,
    $N_{FU}=50$, $\mathbf\tau_{FU}=(0.3, 0.1, 0.03, 0.01)$.
\end{enumerate}

\begin{figure*}[h]
    \centering
    \includegraphics[width=0.99\textwidth]{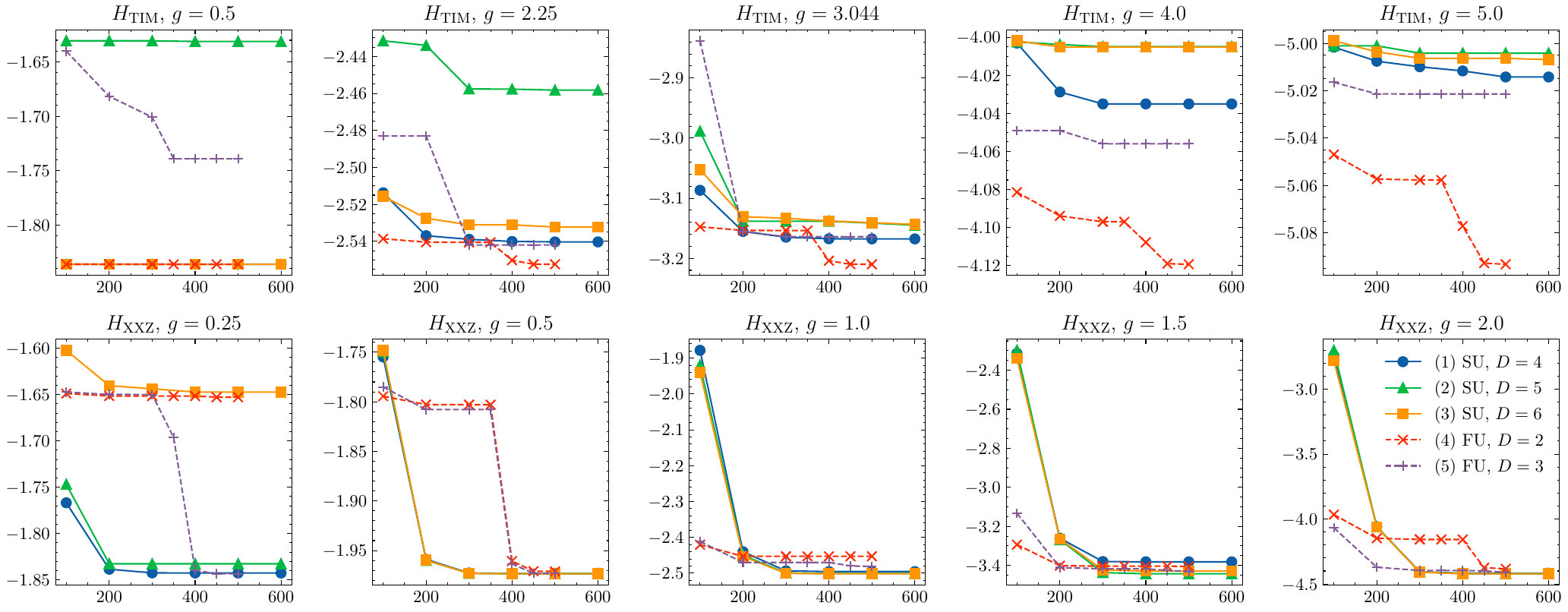}
    \caption{The lowest energies per site found by the PEPS algorithm during ITE for different settings. The x axes are steps and y axes are energies per site. Note that the cost of full update steps are much higher than simple update steps, so the x axes are not uniform in running time.}
    \label{fig:tn-steps-energy}
\end{figure*}

Fig.~\ref{fig:tn-steps-energy} shows the lowest energies found by the PEPS algorithm during ITE for different settings and models. Note that all of the lines are almost flat in the last few points, which indicates that ITE converges and justifies the choice of $N_{SU},N_{FU}$ in these five settings. The lowest energy values found by each variants are shown in Table~\ref{tab:tn-energies}. We observe that Setting 4 (FU, $D=2$) gives the lowest energy for TIM and increasing bond dimension does not necessarily lowers the energy. This might be caused by the truncation error during the contraction, which could be solved by further increasing the contraction bond dimension $\chi$. However, we did not get the chance to verify that due to the limitation computational resources. As for XXZ models, Setting 3 (SU, $D=6$) outperforms other settings in most cases, which implies simple updates approximate full updates better in comparison to TIM. The lowest energy among these five settings for each model is used to compare with other algorithms in this paper.

\begin{table*}[h!]
\centering
\begin{tabular}{|c|c|c|c|c|c|} 
\hline
Models & (1) SU, $D=4$ & (2) SU, $D=5$ & (3) SU, $D=6$ & (4) FU, $D=2$ & (5) FU, $D=3$ \\ [0.5ex] 
\hline\hline
$H_\text{TIM}$, $g={0.5}$ & -1.835865/0.43h & -1.631158/2.29h & \textbf{-1.835905}/10.08h & -1.835902/0.34h & -1.738900/8.07h \\
\hline
$H_\text{TIM}$, $g={2.25}$ & -2.540264/0.45h & -2.458204/2.31h & -2.532226/10.01h & \textbf{-2.552381}/0.58h & -2.541918/7.76h \\
\hline
$H_\text{TIM}$, $g={3.044}$ & -3.167736/0.51h & -3.144920/2.08h & -3.143674/10.09h & \textbf{-3.209989}/0.60h & -3.164132/7.81h \\
\hline
$H_\text{TIM}$, $g={4.0}$ & -4.035041/0.40h & -4.004912/2.23h & -4.005118/10.08h & \textbf{-4.119333}/0.48h & -4.055966/8.84h \\
\hline
$H_\text{TIM}$, $g={5.0}$ & -5.014206/0.46h & -5.004122/2.19h & -5.006872/10.06h & \textbf{-5.093277}/0.53h & -5.021367/11.12h \\
\hline
\hline
$H_\text{XXZ}$, $g={0.25}$ & -1.842531/0.46h & -1.832548/2.20h & -1.647228/10.57h & -1.652827/0.42h & \textbf{-1.843388}/5.96h \\
\hline
$H_\text{XXZ}$, $g={0.5}$ & -1.973211/0.42h & -1.973445/2.10h & \textbf{-1.973462}/10.84h & -1.971098/0.38h & -1.973191/7.33h \\
\hline
$H_\text{XXZ}$, $g={1.0}$ & -2.495433/0.43h & -2.500836/2.07h & \textbf{-2.501652}/10.75h & -2.452904/0.38h & -2.481095/6.24h \\
\hline
$H_\text{XXZ}$, $g={1.5}$ & -3.381753/0.43h & \textbf{-3.442970}/2.06h & -3.428895/10.91h & -3.404844/0.49h & -3.430550/7.16h \\
\hline
$H_\text{XXZ}$, $g={2.0}$ & -4.417899/0.46h & -4.421432/2.07h & \textbf{-4.421658}/11.39h & -4.383387/0.66h & -4.410256/7.28h \\
\hline
\end{tabular}
\caption{Ground state energies per site calculated by PEPS as well as the corresponding running time (the computer system is shared so there might be small fluctuations in running time caused by that). The lowest energy we got for each model is shown in bold face.}
\label{tab:tn-energies}
\end{table*}

\subsection{Neural network quantum states}

    We are interested in the same problem as before, to find the ground state by minimizing the expected value of the energy of our variational state. Instead of defining the quantum state from a parameterized family of probability distributions, now we define a "Neural Quantum State"~\cite{carleo2017solving} (NQS), as the variational Ansatz. In this formulation, a quantum state, $|\psi(\utheta)\rangle$, varies with the interconnections of a neural-network model, $\texttt{NN}(\utheta;\usigma)$,  where $\utheta$ are the weights and biases of the network. 
    The NQS takes a spin-configuration eigenstate as input, $\usigma$, and outputs the associated probability amplitude. 
    \begin{equation}
        \texttt{NN}(\utheta;\usigma) = \langle \usigma | \psi(\utheta)\rangle
    \end{equation}
    
    The energy of the NQS is defined with a normalization term as
    \begin{equation}
        E(\utheta) =  \frac{\langle \psi | \mathcal{H} | \psi \rangle}{\langle \psi | \psi \rangle} = \frac{\sum_{\usigma, \usigma^\prime} \texttt{NN}(\utheta; \usigma) \mathcal{H}_{\usigma, \usigma^\prime} \texttt{NN}(\utheta; \usigma^\prime)}{\sum_{\usigma}|\texttt{NN}(\utheta;\usigma)|^2}
    \end{equation}
    
    Using the completeness relation and $P(\usigma) = |\langle \usigma | \psi(\utheta)\rangle|^2$, find
    \begin{equation}
        E(\utheta) = \sum_{\usigma, \usigma^\prime} P(\usigma) \langle \usigma | \mathcal{H} | \usigma^\prime \rangle \frac{\langle \usigma^\prime | \psi(\utheta)\rangle}{\langle \usigma | \psi(\utheta)\rangle}
    \end{equation}
    which can be estimated using iid samples from the distribution of $P(\usigma)$. This variational Monte Carlo method often uses the Markov chain Metropolis-Hasting algorithm, which generates samples by conditional probabilities of accepting a new sample.
    Finally, the neural network is trained to find the ground state, with a training objective defined like before: $\min_{\utheta} E(\utheta)$ using classical back-propagation techniques.
    
     NetKet~\cite{vicentini2021netket} is an open-source python machine-learning toolbox that implements Quantum Monte Carlo algorithms. Netket is designed to be modular, where the variational state, sampler, optimizer, and Hamiltonian operator are assembled independently before passed to a driver class. Netket creates neural networks using FLAX, a flexible, JAX-based framework. The interface to define variational states comes with pre-built models including FFNN, RBM, ARNN, and Jastrow state functions. Additionally, the sampler module implements several variations of the Monte Carlo sampling functions, which can be reconfigured, for example, by modifying the transition rule in the Metropolis algorthim. Netket supports building lattice-spin models using its Graph and Operator classes. Finally, to perform a ground state search, Stochastic Gradient Descent, a common optimizer choice, estimates the gradient of the Hamiltonian with respect to parameters of the variational state in order to train the VMC state.

    For optimizing the NQS we find that the preconditioned gradient descent method known as Stochastic Reconfiguration (SR) gives a better variational energy than Stochastic Gradient Descent (SGD). For the system size/ number of samples that we considered we did not observe a significant time overhead because of the preconditioning step. 

    For the architecture of NQS we chose a neural net with  a single hidden layer using the $\log(\cosh)$ activation function. The size of the hidden layer was chosen to be half of the number of spins in the quantum system. This was done to ensure that the NQS model has the same number of trainable parameters as the pair-wise AGM Ansatz. The learning rate and diagonal shift parameters for the optimization were both choosen to be $0.01$.

\subsection{Autoregressive Graphical Models}

Each conditional in the AGM Ansatz in \eqref{eq:AGM_pairwise} can be seen as neural net with a single linear layer on $\usigma_{n-1}$, followed by multiplication with $\sigma_i$ and a final sigmoid layer. Then the gradient could be estimated efficiently as in~\eqref{eq:grad_est} with the help of automatic differentiation. So this Ansatz can be implemented as a simple neural net in any of the popular machine learning libraries. The size of the Ansatz is at most $O(n^2)$. For the experiments in this paper, we implemented this method in \textit{PyTorch}. 

Optimization of the VMC objective was performed using the ADAM, which we found to be most effective method.
Methods computationally intensive methods like, Stochastic Reconfiguration (SR) and LMBFGS, did not give good results in comparison to first-order methods during our experiments. Recent work of Wu et al. \cite{wu2022tensor} also observes that ADAM to be more suitable for models using an auto-regressive factorization when compared to SR. On the other hand, for NNQS models the SR preconditioning gives an advantage over first-order methods \cite{vicentini2021netket}.

For the initialization \& hyperparameters setting, we used a fixed Gaussian distribution for setting the initial weights in all the experiments. The hyperparameters involved for each run of the AGM algorithm include number of samples $N$ per step and the schedule of learning rate, the latter of which is determined by the initial learning rate $\alpha_0$ and 1000-step decay rate $\gamma$ (i.e. the learning rate gets multiplied by $\gamma$ every 1000 steps). In our experiments, we fix $N=2^{12}$ and choose $\alpha_0$ and $\gamma$ using hyperparameter optimization for different models.

The hyperparameter optimization step treats the AGM algorithm as a black box and runs it multiple times to determine a relatively good choice of $\alpha_0$ and $\gamma$. In our experiments, we implement this procedure using the \textit{Hyperopt} package~\cite{bergstra2013making}. We empirically choose $\gamma\in[0.8,1.0]$ and $\alpha_0$ in the order of $1\mathrm{E}{-}2$ to $1\mathrm{E}{-}4$ depending on the Hamiltonians. During hyperparameter optimization, the AGM algorithm runs for 10000 iterations with only $2^8$ samples per iteration (instead of $2^{12}$ samples in \figurename \ref{fig:annni} for instance). Also, the AGM algorithm is called at most 30 times to determine each set of hyperparameters. This keeps the time required for hyperparameter optimization lower in comparison to the main experiment while still giving good results. For the disorder averaged experiments, the hyperparameter tuning was done for only one disorder instance at each value of $g.$

The AGM experiments for the unfrustrated TIM and XXZ models were run with two Haswell-E 8-core Xeon CPUs, 64 GiB of DDR4 RAM, and one Nvidia Titan X GPU. For the experiments with D-TIM and ANNNI models, a system with  an Intel Xeon W-11855M Processor, 128 GB DDR4 RAM, and  an A3000 GPU was used.


\section{Additional numerical results}
\label{app:TIM_results}

\begin{figure*}[h]
    \centering
    \includegraphics[width=.99\textwidth]{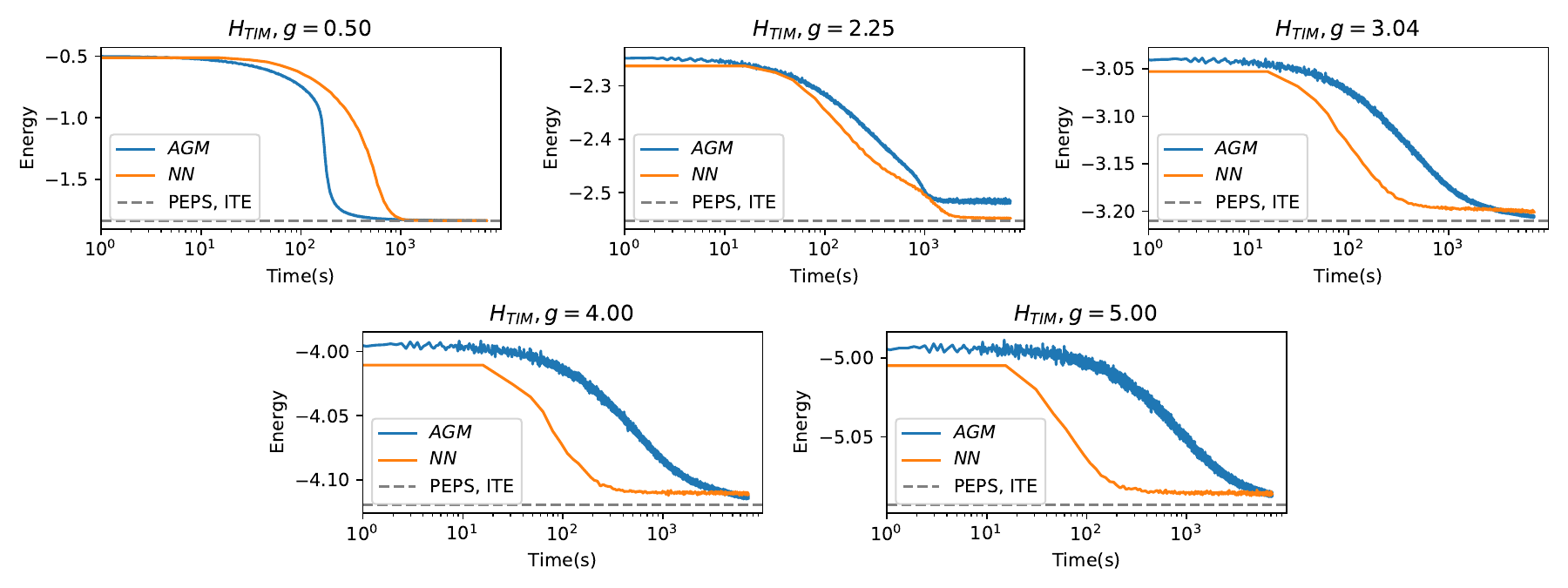}
    \caption{Energy vs Time for TIM model on a $10 \times 10 $ lattice. Simulation performed using $N_s=2^{12}$}
    \label{fig:tim-2d-time-loss-}
\end{figure*}

\begin{figure*}[h]
    \centering
    \includegraphics[width=\textwidth]{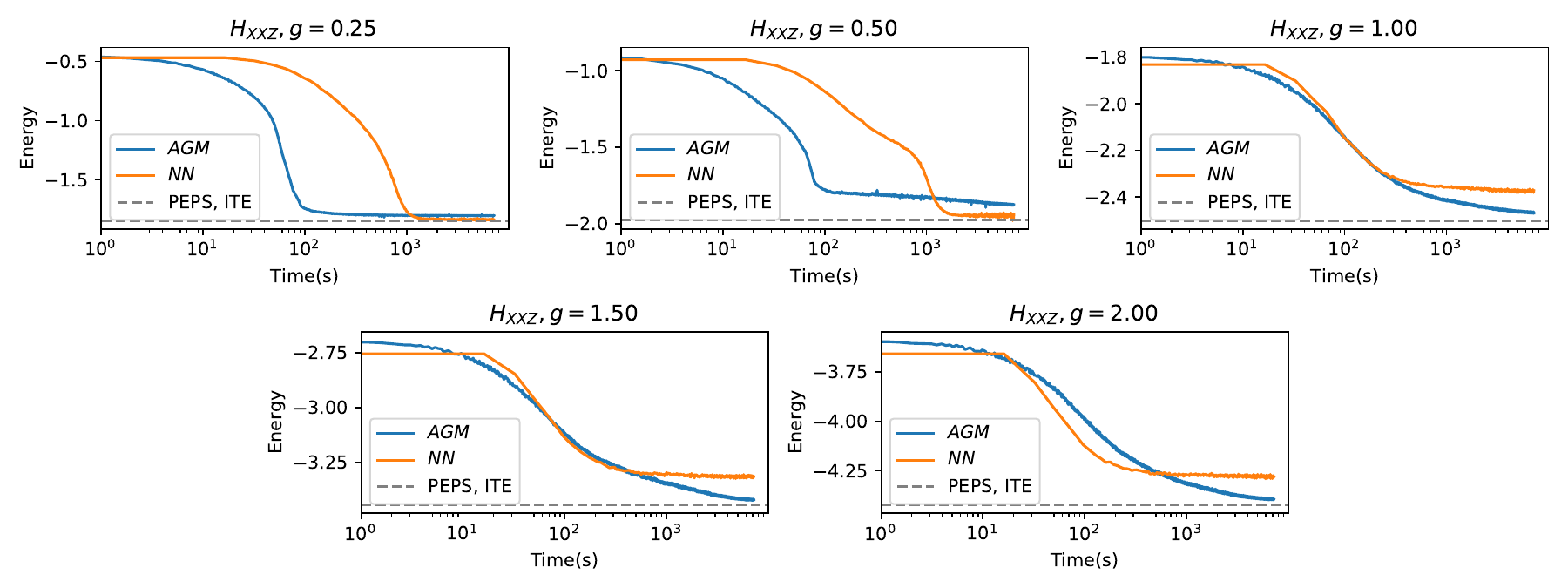}
    \caption{ Energy vs Time for the anti-ferromagnetic XXZ model on a $10 \times 10 $ lattice. Simulation performed using $N_s=2^{12}$}
    \label{fig:axxz-2d-time-loss}
\end{figure*}

The Variational energy plots for $10 \times 10$ TIM and XXZ models for additional values of Hamiltonian parameters are provided in this section for completeness (\figurename \ref{fig:tim-2d-time-loss-} and \figurename \ref{fig:axxz-2d-time-loss}). The number of samples used in these simulations is fixed to $2^{12}$. Overall, we observe a similar performance between AGM and NNQS Ansatze at the end of the variational optimization. As discussed in the main text, we observe that AGM performs better in these simulations in a regime where the off-diagonal terms of the Hamiltonian dominate.

\bibliographystyle{plain}
\bibliography{ref}

\end{document}